%% file: main.tex
\documentclass[aps,pra,superscriptaddress,notitlepage,twocolumn,showpacs]{revtex4-1}
\usepackage[utf8]{inputenc}
\usepackage[usenames,dvipsnames]{xcolor}
\usepackage{amsmath}
\usepackage{amssymb}
\usepackage{mathtools}
\usepackage{esint}
\usepackage{dcolumn}
\usepackage{bm}
\usepackage{bbm}
\usepackage{blindtext}
\usepackage{braket}
\usepackage{natbib}
\usepackage{graphicx}

\usepackage[normalem]{ulem}
\usepackage{enumerate}
\usepackage{xpatch} 

\usepackage{multirow}

\usepackage[hidelinks=true,pdfborder={0 0 0},colorlinks=true,linkcolor=blue,urlcolor=blue,citecolor=blue]{hyperref}

\newcommand{\be}{\begin{eqnarray}}
\newcommand{\ee}{\end{eqnarray}}

\definecolor{dgreen}{rgb}{0.0, 0.5, 0.0}

\renewcommand{\vec}[1]{\mathbf {#1}}

\begin{document}

\title{The building principle of triatomic trilobite Rydberg molecules}

\author{Christian Fey}
\affiliation{Zentrum f\"ur Optische Quantentechnologien, Universit\"at Hamburg,  22761 Hamburg, Germany}
\author{Frederic Hummel}
\affiliation{Zentrum f\"ur Optische Quantentechnologien, Universit\"at Hamburg,  22761 Hamburg, Germany}
\author{Peter Schmelcher}
\affiliation{Zentrum f\"ur Optische Quantentechnologien, Universit\"at Hamburg,  22761 Hamburg, Germany}
\affiliation{The Hamburg Centre for Ultrafast Imaging, Universit\"at Hamburg, Luruper Chaussee 149, 22761 Hamburg, Germany}

\date{\today}
\begin{abstract}
We investigate triatomic molecules that consist of two ground state atoms and a highly excited Rydberg atom, bound at large internuclear distances of thousands of \AA ngstroms. In the molecular state the Rydberg electron is in a superposition of high angular momentum states whose probability densities resemble the form of trilobite fossils. The associated potential energy landscape has an oscillatory shape and supports a rich variety of stable geometries with different bond angles and bond lengths. Based on an electronic structure investigation we analyze the molecular geometry systematically and develop a simple building principle that predicts the triatomic equilibrium configurations.
As a representative example we focus on $^{87}$Rb trimers correlated to the $n=30$ Rydberg state. Using an exact diagonalization scheme
we determine and characterize localized vibrational states in these potential minima with energy spacings on the order of 100 MHz$\times h$.

  \end{abstract}
\maketitle
\section{Introduction}
Ultralong-range Rydberg molecules (ULRM) are a manifestation of a novel type of chemical bond,  where a ground state atom is captured in the electronic cloud of a highly excited Rydberg atom \cite{greene_creation_2000}. In contrast to conventional diatomic molecules, an ULRM possess an oscillatory potential energy surface and huge bond lengths ranging typically from a few hundreds to thousands of Bohr radii $a_0$.
Based on the angular momentum $l$ of the Rydberg electron, two classes of ULRM can be distinguished: Weakly bound non-polar ULRM that correlate to quantum-defect-split Rydberg states with low angular momentum $l< 3$, as well as more deeply bound polar ULRM in which the Rydberg electron is in a superposition of hydrogen-like high-$l$ states and may possess large electric dipole moments on the order of hundreds to thousands of Debye. In allusion to the shape of their electronic probability density, polar ULRM are further subdivided into 'trilobite' molecules with dominant $s$-wave interaction \cite{greene_creation_2000} as well as 'butterfly' molecules \cite{hamilton_shape-resonance-induced_2002, khuskivadze_adiabatic_2002} with dominant $p$-wave interactions. All of these species have been confirmed experimentally \cite{tallant_observation_2012, desalvo_ultra-long-range_2015, booth_production_2015, niederprum_observation_2016, Kleinbach_2017} via one- or two-photon association in ultracold samples of either Rb, Cs or Sr.
Experimental and theoretical research on ULRM demonstrated novel possibilities to tailor molecular properties via weak fields \cite{krupp_alignment_2014, lesanovsky_ultra-long-range_2006,gaj_molecular_2014,niederprum_observation_2016,kurz_electrically_2013, kurz_ultralong-range_2014} and to control atom-atom interactions \cite{Sandor_2017, Thomas_Lippe_Eichert_Ott_2018}. Furthermore, ULRM provide unprecedented access to the physics of 
electron-atom scattering \cite{anderson_photoassociation_2014, sasmannshausen_experimental_2015,
bottcher_observation_2016,schlagmuller_probing_2016, MacLennan_Chen_Raithel_2018} and ion-atom interactions \cite{Kleinbach_2018, Schmid_2018, Engel_2018}.

Having control over the density of the atomic sample and the Rydberg excitation $n$, experiments are able to create and probe not only diatomic ULRM but also polyatomic ULRM. These are bound states between one Rydberg atom and several ground state atoms.  
Although, originally predicted for polar high-$l$ ULRM \cite{liu_polyatomic_2006}, experimental reasearch focussed so far exclusively on non-polar types, that are more easily accessible via one- or two-photon transitions. Experiments with $s$-state ULRM confirmed the existence of few-body states (trimers, tetramers, pentamers) as well as polaronic many-body states, both in excellent agreement with corresponding theoretical models \cite{bendkowsky_rydberg_2010, gaj_molecular_2014, schmidt_mesoscopic_2016, camargo_creation_2018, schmidt_theory_2018}.

From a theoretical as well as from an experimental point of view, the isotropy of the electronic wave function in polyatomic $s$-state ULRM has certain advantages. It simplifies the theoretical models \cite{schmidt_mesoscopic_2016, schmidt_theory_2018} and grants high excitation efficiencies in the experiments, due to comparatively large Franck-Condon factors. An obvious drawback of this isotropy is, however, that there is only weak control over the molecular geometry, especially the angular geometry. This is different for polyatomic $p$- and $d$-state ULRM \cite{fey_stretching_2016, Fey_2018}, which exist in linear and bent geometries. The Hilbert space of energetically available electronic states is here larger, and provides more possibilities for the Rydberg electron to optimize its wave function. Consequently, for polyatomic trilobite ULRM, with their large manifold of energetically degenerate hydrogenic states, even more complex geometries with deeper potential wells are expected. Previous studies predicted exotic properties of these molecules, such as their capability to form Borromean like states\cite{liu_ultra-long-range_2009} or the appearance of quantum scars \cite{Luukko_Rost_2017}. Furthermore the large electric dipole moments allow to tune their geometry via weak electric fields \cite{Fernandez_2016}. However, all of these theoretical works focused so far only on constrained geometries, such as linear,  planar or cubic configurations, or on polyatomic trilobite states in random configurations \cite{Luukko_Rost_2017}. Consequently, a thorough understanding of the molecular geometry of even the simplest polyatomic trilobite ULRM, the trilobite trimer, is missing. In our work we aim at closing this gap. We analyze the full three-dimensional potential energy landscape of trilobite trimers by means of an investigation of their electronic structure and derive a simple building principle that explains the resulting equilibrium positions. Subsequently we employ an exact diagonalization scheme to predict energies and wave functions of bound vibrational states, which are relevant for spectroscopic measurements. 

This work is organized as follows. In Section II we present the electronic Hamiltonian of the molecular system and derive the corresponding potential energy surfaces (PES). Furthermore we identify equilibrium positions and explain their geometrical arrangement. In Section III we provide the theoretical framework for the description of vibrational states in these PES. Subsequently, we 
present energies and probability densities of vibrational states and discuss their properties. Section IV contains our conclusions.

\section{energy landscape}

\begin{figure}[h]
\includegraphics[width=0.8\linewidth]{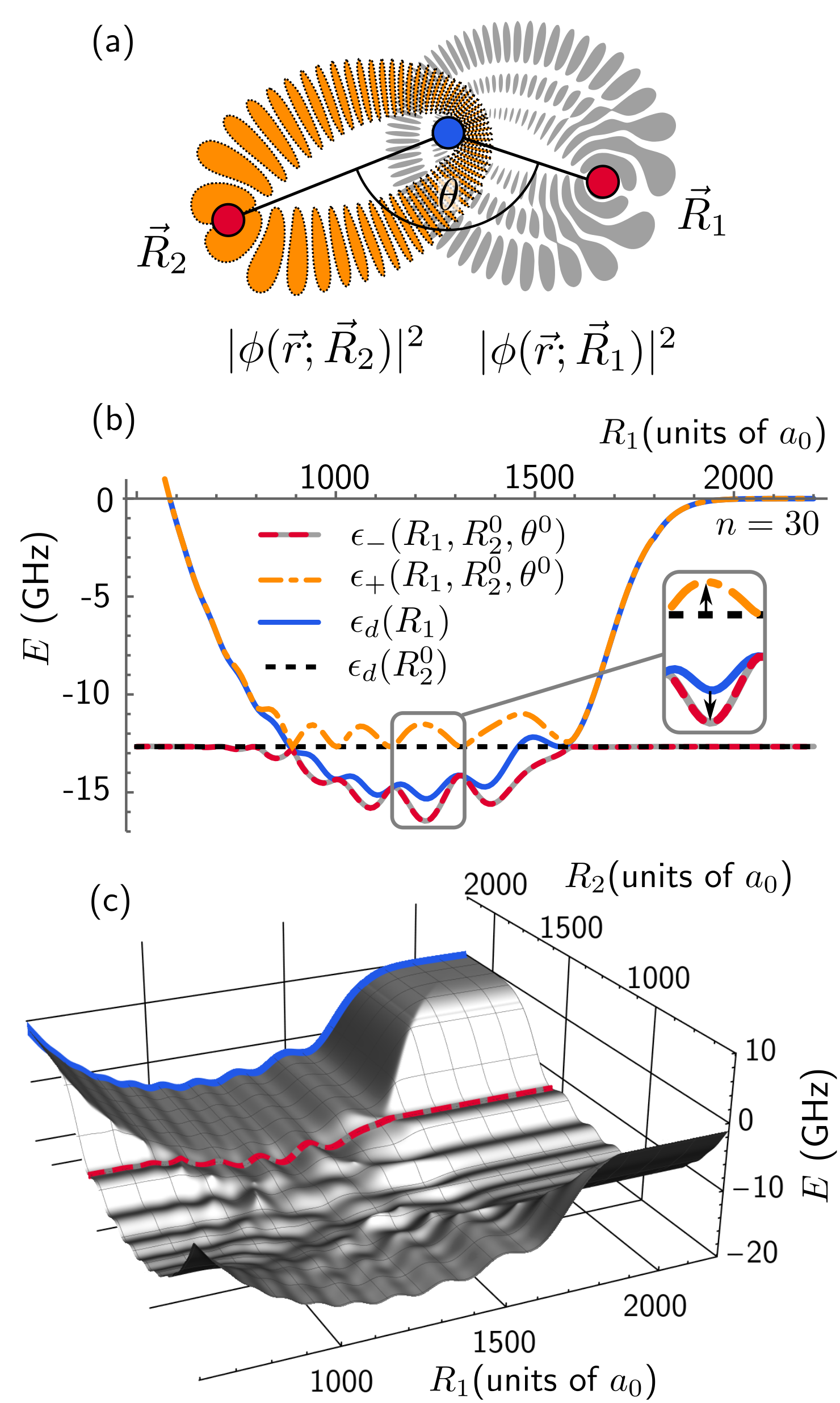}
\caption{(a) The Trilobite trimer consists of two ground state atoms (red) at positions $\vec{R}_{1/2}$ relative to the ionic core (blue). The Rydberg electron (here $n=30$) is in a superposition of the two trilobite states $\phi(\vec{r};\vec{R}_{1/2})$ (orange vs. gray density). (b) Cuts of the two trimer PES $\epsilon_{\pm}(R_1,R_2^0,\theta^0)$ for fixed $R_2=R_2^0=1563$~$a_0$ and $\theta=\theta^0=0.12 \pi$. The energy of the hydrogenic states with $n=30$ is set to zero.  These potentials are compared to the diatomic PES  $\epsilon_d(R_1)$  and  $\epsilon_d(R_2^0)$. A magnification of the deviations is presented in the inset. (c) 2D cut of the lower PES $\epsilon_{-}(R_1,R_2,\theta^0)$ where only $\theta^0=0.12 \pi$ is fixed. The colored lines mark cuts $\epsilon_{-}(R_1,R_2^0,\theta^0)$ (red-gray dashed) and $\epsilon_{d}(R_1)$ (solid blue) that are also visible in (b).    }
\label{fig:sketch_pes}
\end{figure}

A general polyatomic ULRM consists of an ionic core (here at the coordinate origin), a Rydberg electron at position $\vec{r}$  and $N$ neutral ground-state atoms at positions $\vec{R}_i$ where $i=1,\dots,N$. A sketch of the setup for $N=2$ is presented in Fig.~\ref{fig:sketch_pes}~(a). In the Born-Oppenheimer approximation the adiabatic electronic Hamiltonian is given by $H=H_0 + V$ where $H_0$ describes the Rydberg electron in its ionic core potential while $V$ is the interaction between the Rydberg electron and the ground state atoms.
In dependence of the electronic angular momentum $l$, the eigenstates of $H_0$ can be divided into low-$l$ and high-$l$ states. Due to their centrifugal barrier, high-$l$ states (typically $l\geq3$) are shielded from the ionic core. To a good approximation they are given by hydrogen wave functions $\varphi_{nlm}(\vec{r})$ with energies $-1/(2n^2)$ (in atomic units) where  $n$ and $m$ are the principal and the magnetic quantum number, respectively. 
However, due to the presence of the ground state atoms (perturbers) inside the Rydberg orbit, the hydrogenic states become coupled. 
We focus on $^{87}$Rb ULRM where this coupling is small compared to the energy splitting between high-$l$ ($l\geq3$) and low-$l$ ($l<3$) states, such that a perturbative approach is appropriate \cite{greene_creation_2000}.
Working in atomic units, we model this interaction with the perturber via a contact potential \cite{fermi_sopra_1934, omont_theory_1977, greene_creation_2000}
\begin{equation}
V= \sum_{j=1}^N 2 \pi  a[k(R_j )] \delta(\vec{r}-\vec{R}_j).
\label{eqn:pseudopotential}
\end{equation} 
The energy dependence of the scattering length is obtained via modified effective range theory \cite{spruch_modification_1960, omalley_modification_1961} $a(k)=a(0)+(\pi/3) \alpha k$ with the electron wavenumber $k$, the Rb(5$s$) polarizability $\alpha=319.2$ and the zero-energy scattering length $a(0)=-16.1$ for $e$-Rb(5s) triplet scattering ($^3S$) \cite{bahrim_3se_2001}.
In a semi-classical approximation the wave wave number is determined via $k^2/2-1/R=-1/(2n_0^2)$, where $n_0$ is the principal quantum number of interest. 
Despite its simplicity, the Hamiltonian $H$ captures the essential features of trilobite ULRM. Quantitative corrections originate from the $^{87}$Rb fine and hyperfine structure, additional $p$-wave interactions as well as spin-spin and spin-orbit couplings  \cite{khuskivadze_adiabatic_2002, hamilton_shape-resonance-induced_2002, anderson_photoassociation_2014, eiles_hamiltonian_2017,eiles_ultracold_2016,hummel_2018,hummel_2018_s_state}. Furthermore there exist non-perturbative methods relying on Greens's function methods \cite{khuskivadze_adiabatic_2002,bendkowsky_rydberg_2010, fey_comparative_2015, tarana_adiabatic_2016}.

For dimers ($N=1$, $\vec{R}_1=\vec{R}$) the contact interaction gives rise to an electronic eigenstate that strongly localizes on the perturber and resembles the shape of a trilobite fossil, see Fig. \ref{fig:sketch_pes} (a) for two examples. Performing first order perturbation theory in the Hilbert space of quasi-degenerate hydrogenic states with $n=n_0$ and $l\geq3$, its wave function can be expressed as $\Psi(\vec{r};\vec{R}) = \mathcal{N}\phi(\vec{r};\vec{R})$ with the trilobite orbital
\begin{equation}
\phi({\vec{r};\vec{R}})=  \sum \limits_{l=3}^{n_0-1} \sum \limits_{m=-l}^{l} \varphi^{*}_{n_0lm}(\vec{R})\varphi_{n_0lm}(\vec{r}) \ .
\label{eqn:trilobite}
\end{equation}
and the normalization constant $\mathcal{N}=\phi(\vec{R};\vec{R})^{-1/2}$ \cite{liu_polyatomic_2006}.
It is the superposition of all hydrogen states that minimizes its energy by maximizing its density on the perturber. This density is azimuthally symmetric around the internuclear axis $\vec{R}$. In particular, if $\vec{R}$ points along the $z$-axis, only terms with $m=0$ contribute. The associated energy shift (first order) is given by
\begin{equation}
\epsilon_d(R)= 2 \pi a [k(R)] \phi(\vec{R};\vec{R}).
\label{eqn:dimer}
\end{equation}
This is the potential energy surface (PES) of the molecule.
It is straightforward to show that the PES depends only on $R=|\vec{R}|$. An example for $n_0=30$ (solid blue line) is presented in Fig.~\ref{fig:sketch_pes}~(b). The deepest minima support a series of localized vibrational states \cite{greene_creation_2000}.

The electronic structure becomes altered when a second perturber is present ($N=2$).
To obtain the electronic trimer states efficiently, we separate the Hilbert space with $n=n_0$ and $l\geq 3$ into the subspace spanned by the two dimer solutions $\phi({\vec{r};\vec{R}_j})$ with $j=1,2$  and the remaining complement, that may be obtained via Gram-Schmidt orthogonalization. One can show that all states in the complement have nodes at $\vec{R}_1$ and $\vec{R}_2$ and do thus not probe the ground state atoms \cite{liu_ultra-long-range_2009}. A proof is provided in Appendix A. 
Consequently, within first order perturbation theory, we can express the trimer state as a linear combination of the two dimer solutions \cite{liu_polyatomic_2006, liu_ultra-long-range_2009,eiles_ultracold_2016,fey_stretching_2016}  
\begin{equation}
\psi({\vec{r};\vec{R}_1,\vec{R}_2})=  \sum \limits_{j=1}^2 c_j \phi({\vec{r};\vec{R}_j})
\label{eqn:dimer_ansatz}
\end{equation}
with coefficients $c_j$ that depend on $\vec{R}_1$ and $\vec{R}_2$. This situation is visualized in Fig. \ref{fig:sketch_pes} (a) schematically, where we take into account, that the shape of the trilobite state $\phi({\vec{r};\vec{R}_j})$ depends explicitly on the nuclear coordinate $\vec{R}_j$. 
Eigenstates of this two-level system are determined by solving the corresponding generalized eigenvalue problem for $H$. The two resulting PES 
\begin{equation}
\begin{split}
\epsilon_\pm (\vec{R}_1,\vec{R}_2)&= \frac{\epsilon_d(R_1)+\epsilon_d(R_2)}{2} \\
& \pm \frac{1}{2} \sqrt{\left(\epsilon_d(R_1)-\epsilon_d(R_2)\right)^2 +4 c(\vec{R}_1,\vec{R}_2)},
\end{split}
\label{eqn:trimer_energy_first_order}
\end{equation}
can be expressed in terms of the diatomic potentials (\ref{eqn:dimer}) and a term
\begin{equation}
 c(\vec{R}_1,\vec{R}_2)= 4 \pi^2 a[k(R_1)] a[k(R_2)] \left|\phi(\vec{R}_2;\vec{R}_1)\right|^2.
 \label{eqn:coupling}
 \end{equation}
The latter contains the trilobite orbital (\ref{eqn:trilobite}) as a function of the two nuclear coordinates. 
It depends in addition to $R_1=|\vec{R}_1|$ and $R_2=|\vec{R}_2|$, also on the relative angle $\theta=\arccos \left[(\vec{R}_1 \cdot \vec{R}_2)/(R_1 R_2)\right]$ and adds thus an anisotropy to the PES. Furthermore it satisfies $c(\vec{R}_1,\vec{R}_2)=c(\vec{R}_2,\vec{R}_1)$. 

Exemplary cuts of the PES $\epsilon_\pm(R_1,R_2^0,\theta^0)$ are presented in Fig.~\ref{fig:sketch_pes} (b) for fixed $R_2=R_2^0=1563\, a_0$ and $\theta=\theta^0=0.12 \pi$ but variable $R_1$ (dashed gray-red and dashed-dotted yellow line).
These potentials are compared to the corresponding diatomic PES $\epsilon_d(R_1)$ and $\epsilon_d(R^0_2)$ (solid blue and dashed black line), i.e. the PES when the presence of the second ground state atom is ignored.
While the trimer PES coincide with the dimer PES for very large and very small separations $R_1$, there is an intermediate regime, here $900 \,a_0<R_1<1600 \, a_0$, where one finds substantial deviations.
These deviations result solely from the term  $c(\vec{R}_1,\vec{R}_2)$, which, based on the structure of ($\ref{eqn:trimer_energy_first_order}$), can be interpreted as an effective coupling of the two dimer states $\phi(\vec{r};\vec{R}_1)$ and $\phi(\vec{r};\vec{R}_2)$.
E.g. whenever one has $c(\vec{R}_1,\vec{R}_2) = 0$, there is no coupling and the two PES $\epsilon_\pm(R_1,R_2^0,\theta^0)$ coincide with the diatomic PES. In this limit one has
\begin{align}
 \epsilon_+(R_1,R_2,\theta)&=\max(\epsilon_d(R_1),\epsilon_d(R_2)) \nonumber \\
  \epsilon_-(R_1,R_2,\theta)&=\min(\epsilon_d(R_1),\epsilon_d(R_2)) .
  \label{eqn:no_coupling}
 \end{align} 
In contrast, a non-vanishing coupling $c(\vec{R}_1,\vec{R}_2)$ introduces a level repulsion between the two diatomic PES. This is represented by the black arrows in the inset in Fig.~\ref{fig:sketch_pes}~(b).
Importantly, in the lower PES $\epsilon_-(R_1,R_2,\theta)$ this effect leads to an energy drop below the dimer PES and can therefore stabilize trimer states. This effects is also visible in the higher-dimensional cut of the PES $\epsilon_-(R_1,R_2,\theta^0)$ in Fig.~ \ref{fig:sketch_pes}~(c). Level repulsion takes place in the region where $R_1<1600 a_0$ and $R_2<1600 a_0$, where it induces a rich oscillatory pattern with many radial minima, that are energetically well below the dimer PES (solid blue line).
To provide some visual orientation, the dashed red line marks the curve $\epsilon_-(R_1,R_2^0,\theta^0)$ and links Fig.~\ref{fig:sketch_pes}~(b) to Fig.~\ref{fig:sketch_pes}~(c).

\begin{figure*}[t]
\includegraphics[width=0.9\linewidth]{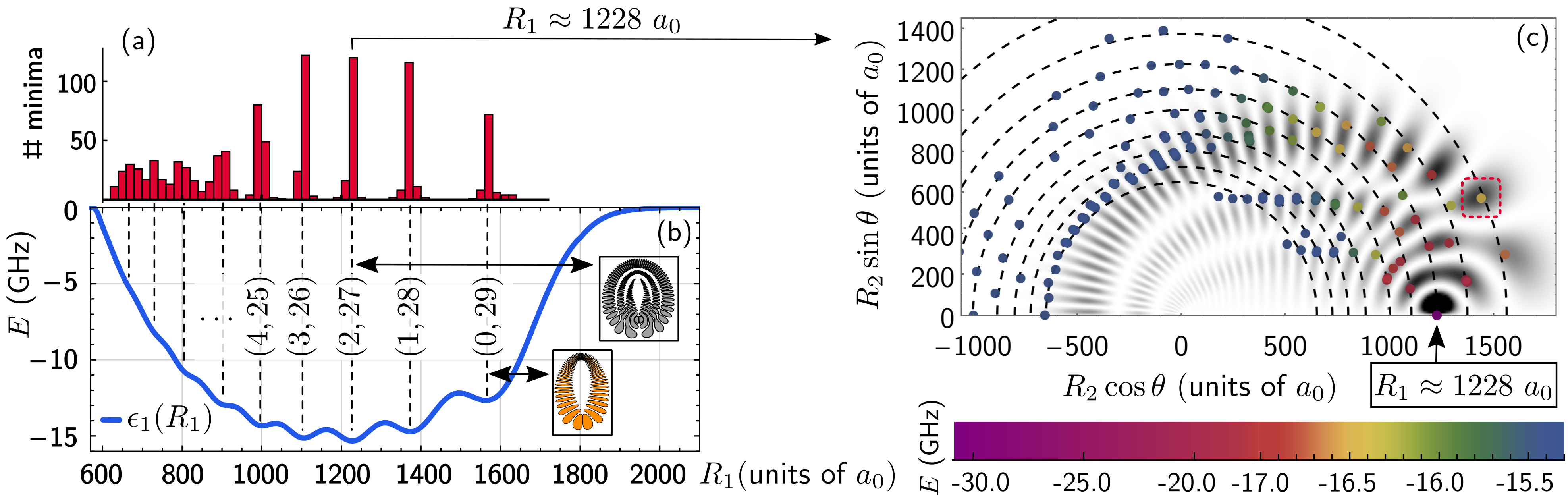}
\caption{Analysis of the positions of minima $(R_1,R_2,\theta)$ in the trilobite trimer PES $\epsilon_-(R_1,R_2,\theta)$ given in (\ref{eqn:trimer_energy_first_order}). (a) The histogram depicts the number of minima as a function of their coordinate $R_1$. (b) Peak positions $R_c$ in the histogram (dashed lines) occur at radii $R_1$ where the trilobite wave function $\phi(\vec{r};\vec{R}_1)$ is dominated by orbitals having integer number of nodes $(n_1,n_2)$ along two different elliptical directions \cite{Granger_2001}. The insets present exemplary probability densities for $(0,29)$ and $(2,27)$. Minima in the diatomic PES $\epsilon_d(R_1)$ (blue line) coincide with these radii $R_c$. (c) Each dot represents the coordinates $R_2$ and $\theta$ of a minimum in the subset with $R_1\approx 1228\, a_0$. The color encodes the potential well depths. The additional gray shading represents the electronic density of the trilobite dimer $|\phi(\vec{R}_2;\vec{R}_1)|^2$ associated to the subset with $R_1=1228 a_0$ and characterized by $(2,27)$. Deep minima occur at positions where circles of radius $R_c$ (dashed lines) intersect the peaks of the trilobite density, e.g. at the position marked by the dashed red square. Bound vibrational states in this minimum
are presented in Fig.~\ref{fig:vibrational_states} as an example case.}
\label{fig:dimer_histogram}
\end{figure*}

The exemplary cuts of the PES in Fig.~\ref{fig:sketch_pes}~(b) and (c) demonstrate that the coupling $c(\vec{R}_1,\vec{R}_2)$ has a crucial impact on the PES. In the following we study as to which extent this mechanism affects the equilibrium configuration of the trimer (stable in $R_1$, $R_2$ and $\theta$). To this aim we evaluate $\epsilon_-(R_1,R_2,\theta)$ on a cubic grid and detect all local minima. Surprisingly, this yields a large set of the order of thousand equilibrium positions.   
We analyze these positions in two steps.
Firstly, we classify the equilibria with respect to the coordinate $R_1$. Subsequently, in a second step, we focus on the structure with respect to the remaining coordinates $R_2$ and $\theta$.  

The histogram in Fig. \ref{fig:dimer_histogram}~(a) presents the abundance of minima in dependence of the coordinate $R_1$. Due to the indistinguishability of the two ground state atoms, the histogram does not change if one replaces $R_1$ by $R_2$. As can be seen, the minima are not distributed homogeneously along the $R_1$ axis but cluster around certain separations  $R_c \in \{650, 725, 802, 886,1001,1105,1228,1374,1563\}\, a_0$ marked by vertical dashed lines in Fig.~\ref{fig:dimer_histogram}~(b).
Comparing these values to the shape of the dimer PES in \ref{fig:dimer_histogram}~(b), we find that the positions of the strongest peaks in the histogram coincide with the equilibrium positions of the dimer PES. 
Moreover, all peak positions can be identified with the critical radii found in \cite{Granger_2001} at which the trilobite state $\phi(\vec{r};\vec{R})$ satisfies  semiclassical Einstein-Brillouin-Keller quantization conditions.
The elliptically shaped densities of these states can be characterized by two integers $(n_1,n_2)$ counting the nodes along different elliptical directions. Exemplary probability densities for $(0,29)$ and $(2,27)$ are depicted in Fig.~\ref{fig:dimer_histogram}. We interpret this results in the following way: The radial structure of trimer PES is governed dominantly by the dimer PES. For instance the cut $\epsilon(R_1,R_2^0,\theta^0)$ in Fig.~\ref{fig:sketch_pes}~(dashed gray-red line) is on a large scale well approximated by (\ref{eqn:no_coupling}) (solid blue and dashed black line). However, the coupling $c(\vec{R}_1,\vec{R}_2)$ induces oscillatory deviations and leads to a substructure which is not captured by (\ref{eqn:no_coupling}). Due to these deviations the peaks in the histogram are not sharp but possess a certain width. Furthermore, when Einstein-Brillouin-Keller quantization conditions are fulfilled, these deviations are sufficiently strong to induce minima at radii which are not stable in the diatomic system, e.g. at $R_1= 800 \, a_0$. The formation of ultralong-range Rydberg trimers with repulsive two-body interaction studied in  \cite{liu_ultra-long-range_2009} is a special example for this effect.     
       
Having analyzed the clustering with respect to the $R_1$ coordinate we focus, in a second step, on the configuration of the remaining coordinates $R_2$ and $\theta$. To this aim we select subsets of minima sharing the same $R_1\approx R_c$, i.e. belonging to the same cluster.
As an example Fig.~\ref{fig:dimer_histogram}~(c) depicts the minima of the cluster with $R_1 \approx 1228 \, a_0$. Every dot represents the coordinates of a minimum, i.e. $R_2$ and $\theta$, in the plane perpendicular to $\vec{R}_1$.
The dot color encodes the well depth ranging from -30 GHz to -15 GHz.
In addition to the minima we present the trilobite density from (\ref{eqn:coupling}) $|\phi(\vec{R}_2;\vec{R}_1)|^2$ (shaded gray) as well as the regions where $|R_2|$ coincides with cluster radii $R_c$ by dashed circles.
Minima are expected to support stable trimer states only if their depth is significantly lower than the depth of the dimer PES $\epsilon_d(1228 \, a_0)\approx -15$ GHz.
As can be seen in Fig.~\ref{fig:dimer_histogram}~(c), sufficiently deep minima (with color coding different from blue) occur always close to positions where the dashed circles intersect with strong peaks in the trilobite density. An example for such an intersection is marked by the dashed red square. This geometrical arrangement of the minima is a consequence of the coupling in (\ref{eqn:coupling}).
While radial coordinates of the minima are mostly determined by the properties of the diatomic system via Eqn. (\ref{eqn:no_coupling}), the angular structure is dominated by the coupling (\ref{eqn:coupling}) proportional to the trilobite density. This can be formulated as a simple bottom-up building principle: Based on an existing stable trilobite dimer, a stable trilobite trimer can be constructed by placing an additional ground state atom in a peak of the diatomic trilobite density (shaded gray density in Fig. \ref{fig:dimer_histogram}).
Since the trilobite density possesses many peaks at different positions, this binding mechanism gives rise to a plethora of equilibrium geometries.
The crucial role of diatomic trilobite orbitals as building blocks for polyatomic ULRM has also been pointed out in \cite{liu_polyatomic_2006,liu_ultra-long-range_2009,eiles_ultracold_2016,Luukko_Rost_2017}. These studies focused, however, on constrained geometries or random configurations of the molecules and did not capture its angular equilibrium structure. %
These angular structures result from the mixing of Rydberg wave functions with different quantum numbers $l$ and $m$ as described in (\ref{eqn:dimer_ansatz}). Bent equilibrium geometries are therefore absent in polyatomic $s$-state ULRM with almost isotropic PES but can occur also in $p$- and $d$-state trimers \cite{bendkowsky_rydberg_2010, gaj_molecular_2014,schmidt_mesoscopic_2016,fey_stretching_2016,Fey_2018} 
\section{vibrational states}

In the following we focus on one of the trimer equilibrium positions in more detail and predict the supported vibrational states. This is the minimum marked by the dashed red line in Fig. \ref{fig:dimer_histogram} (c) with coordinates $(R_1^0,R_2^0,\theta^0)=(1228\,a_0,1563 \, a_0,0.12 \pi)$. This analysis serves as an example case to illustrate properties of bound states that occur also in other minima of the PES. 

After separating the center-of-mass motion, the Hamiltonian for the relative nuclear motion can be written as $H^\text{rel}=H^\text{vib}+H^\text{rovib}$, where $H^\text{vib}$ describes pure vibrational dynamics (depending only on $R_1$, $R_2$, $\theta$) and $H^\text{rovib}$ describes rotational as well as rovibrational dynamics. The vibrational part reads \cite{handy_variational_1982,handy_derivation_1987,fey_stretching_2016}
 
\begin{align}
H^\text{vib}&= \frac{1}{m}\left[-\frac{\partial^2}{\partial R_1^2}-\frac{\partial^2}{\partial R_2^2}- \cos \theta \frac{\partial}{\partial R_1}\frac{\partial}{\partial R_2}  \right] \nonumber \\
&- \frac{1}{m} \left(\frac{1}{R_1^2}+ \frac{1}{R_2^2}- \frac{\cos\theta}{R_1 R_2} \right)\left(\frac{\partial^2}{\partial \theta^2} + \cot \theta \frac{\partial}{\partial \theta}  \right) \nonumber \\
&-\frac{1}{m}\left(\frac{1}{R_1 R_2}-\frac{1}{R_2}\frac{\partial}{\partial R_1}
-\frac{1}{R_1}\frac{\partial}{ \partial R_2}
\right)\left(\cos \theta + \sin \theta \frac{\partial}{\partial \theta} \right) \nonumber\\
&+ \epsilon_-(R_1,R_2,\theta) \ .
\label{eqn:Hamiltonian_nuclear}
\end{align}
This Hamiltonian acts on wave functions $\chi(R_1,R_2,\theta)$ being normalized as $\int dR_1 dR_2 d \theta \sin \theta |\chi(R_1,R_2,\theta)|^2=1$.
The total angular momentum $J$ of the nuclei is conserved and we focus on $J=0$, for which case the rovibrational part of the Hamiltonian vanishes.
For $^{87}$Rb we use $m=1.58 \cdot 10^5$ a.u. and consider only bosonic states which satisfy, according to spin-statistics,  $\chi(R_1,R_2,\theta)=\chi(R_2,R_1,\theta)$.

\begin{figure}[t]
\includegraphics[width=0.99\linewidth]{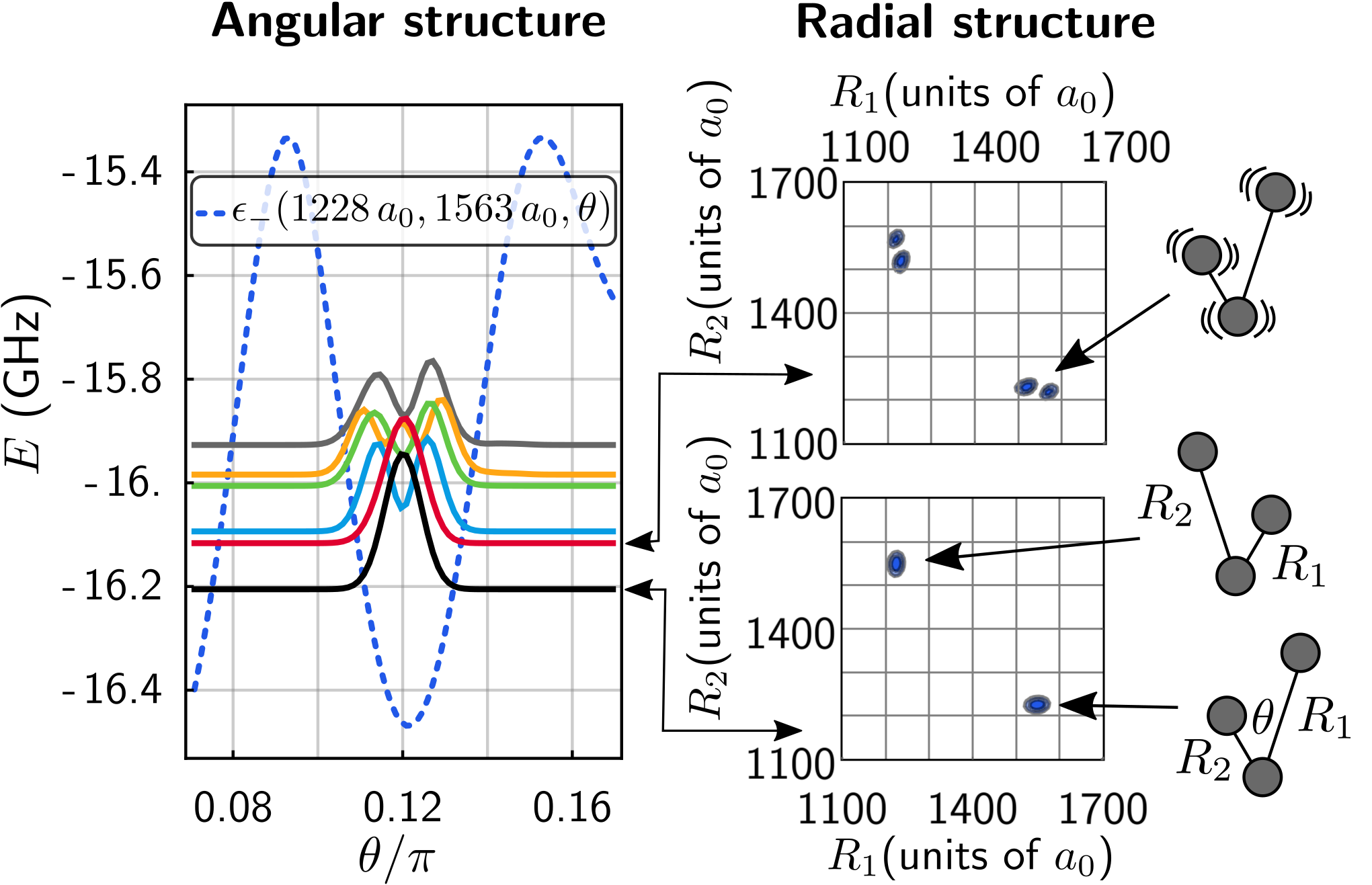}
\caption{Bound vibrational states localized in the minimum $(R_1^0,R_2^0,\theta^0)=(1228\,a_0,1563 \, a_0,0.12 \pi)$, that is marked by the dashed red line in Fig. \ref{fig:dimer_histogram}. (left) Angular cuts through the potential minimum (dashed blue lines) and reduced angular densities of vibrational states (solid lines). The offset is adjusted to their binding energy. (right) Contour plots in the $R_1$-$R_2$ plane of reduced radial densities for the two lowest states. Pictorial representation of the molecular geometry are used to interpret these densities.}
\label{fig:vibrational_states}
\end{figure}

Eigenstates of (\ref{eqn:Hamiltonian_nuclear}) are obtained numerically in position space representation on a three-dimensional cubic grid. For the $R_1$ and $R_2$ direction we use equidistant grid points and build the derivative operators via finite difference expressions. For the $\theta$ degree of freedom we employ a discrete variable representation (DVR) approach \cite{Beck_Meyer_2000}. In a first step Legendre polynomials $P_l(\cos \theta)$ are used as basis functions to construct all kinetic energy operators related to $\theta$ as well as the operator $\cos(\theta)$. In a second step these operators are transformed into a new basis of eigenstates of $\cos(\theta)$, which can be viewed as a discrete approximation of position states $\ket{\theta}$. This approach is non-variational but has the advantage that the operator of the PES $\epsilon_-(R_1,R_2,\theta)$ is diagonal, i.e. there is no need to evaluate overlap integrals. We achieve good convergence by using typically a set of $80$ gridpoints in each dimension.

Fig. \ref{fig:vibrational_states} presents reduced densities of the six energetically lowest vibrational states $\chi(R_1,R_2,\theta)$ that localize in the potential well at $(R_1^0,R_2^0,\theta^0)=(1228\,a_0,1563 \, a_0,0.12 \pi)$.
Out of all resulting eigenstates, the states shown in Fig.~\ref{fig:vibrational_states} were obtained by selecting those having the largest integrated probability density in the potential well considered.
The angular densities ($\int dR_1 dR_2|\chi(R_1,R_2,\theta)|^2$) of these states are depicted in Fig.~\ref{fig:vibrational_states}~(left) together with an angular cut of the PES through the minimum $\epsilon_-(1228 \, a_0,1563 \, a_0,\theta)$.
All depicted densities are very localized around the minimum $\theta^0$.  The ground state has an energy of -16.2 GHz and a Gaussian-shaped angular density.
The corresponding radial density ($\int d\theta |\chi(R_1,R_2,\theta)|^2 \sin \theta$) is presented in Fig.~\ref{fig:vibrational_states}~(right) and exhibits two peaks. One near $(R_1,R_2)=(R_1^0,R_2^0)$ the other near $(R_1,R_2)=(R_2^0,R_1^0)$. This double-peak structure is a consequence of the bosonic symmetry and implies that the molecule is in a superposition of the two 'check-mark' geometries, that are shown as cartoons in Fig.  \ref{fig:vibrational_states}.           
The next higher vibrational states appears approximately 100 MHz above the ground state. While its angular density resembles the density of the ground state, its radial density possesses additional nodes that indicate an excited stretching mode vibration, see pictorial representation in Fig. \ref{fig:vibrational_states}. Even higher excited states exhibit combined bending and stretching excitations.
Altogether, the resulting vibrational states demonstrate that the bent equilibrium geometries of trilobite trimers are stable enough to support a number of vibrational states and that their vibrational spacing is large enough to be resolved in current experimental setups \cite{Kleinbach_2017}.

\section{Conclusions}
We presented the rich molecular structure of trilobite trimers and derived a simple building principle that explains their geometry. Starting from a stable diatomic trilobite , a robust trimer can be formed when the second ground state atom is placed in a density maximum of the diatomic trilobite wave function. For $^{87}$Rb trimers correlated to $n=30$, we demonstrated that the resulting potential minima are deep enough to support a series of localized vibrational states with energy spacings on the order of 100 MHz.
The plethora of equilibrium geometries opens fascinating possibilities to control the arrangement of triatomic molecules but poses also a new challenge to experiments: States in different molecular geometries can have comparable vibrational energies and can therefore hardly be distinguished spectroscopically. Future studies might therefore investigate as to which extent external fields or optical lattices could serve as additional selection mechanisms. Furthermore it might be interesting to generalize the derived building principle of trilobite trimers to tetramers or even larger clusters.

\subsection*{Acknowledgments}
F.H. and P.S. acknowledge support from the German
Research Foundation within the priority program "Giant Interactions in Rydberg Systems" (DFG SPP 1929
GiRyd). C.F. gratefully acknowledges a scholarship by
the Studienstiftung des deutschen Volkes. We thank M. T. Eiles for fruitful discussions.

\input{main.bbl}


\appendix
\section{trilobite orbitals}
To derive and to interpret the PES of the triatomic molecule (\ref{eqn:trimer_energy_first_order}) we employed a two-dimensional basis that consists of the trilobite orbitals $\phi(\vec{r};\vec{R}_1)$ and $\phi(\vec{r};\vec{R}_2)$, see (\ref{eqn:dimer_ansatz}). One can show, that this approach is exact in the sense, that it yields the same PES (\ref{eqn:trimer_energy_first_order}) and the same electronic states as the diagonalization of the interaction $V$ in the larger basis set of hydrogenic states $\varphi_{n_0lm}(\vec{r})$ with fixed $n_0$, $3\leq l\leq n_0-1$ and $|m|\leq l$.
We proof this by demonstrating that the interaction matrix $V$ in (\ref{eqn:pseudopotential}) can be written as
\begin{equation}
V=\sum_{j=1}^N 2 \pi a[k(R_j)]  \ket{\phi_j} \bra{\phi_j} , 
\label{eqn:pseudo_appendix}
\end{equation}
where $\ket{\phi_j}$ denotes the unormalized trilobite state 
with wave function  $\braket{\vec{r}|\phi_j}=\phi(\vec{r};\vec{R}_j)$ as defined in (\ref{eqn:trilobite}). The number of ground state atoms $N$ is in our case $N=2$. From (\ref{eqn:pseudo_appendix}) it becomes evident that all basis states which are perpendicular to the trilobite orbitals $\phi(\vec{r};\vec{R}_1)$ and $\phi(\vec{r};\vec{R}_2)$ will not interact with the ground state atoms and will, therefore, not contribute to the trimer PES, nor to its molecular states.

To proof (\ref{eqn:pseudo_appendix}) we introduce the multiindex $\alpha=(n_0,l,m)$ that labels all basis states compactly as $\varphi_\alpha(\vec{r})$.  The matrix elements of the delta potential of the $j$-th perturber read in this basis 
\begin{equation}
\bra{\varphi_\alpha} \delta(\vec{r}-\vec{R}_j)\ket{\varphi_{\alpha'}}= \varphi_\alpha^*(\vec{R}_j) \varphi_{\alpha'}(\vec{R}_j).
\label{eqn:delta_appendix}
\end{equation}  
A particular property of this matrix is that all rows are linear dependent, e.g. the first row 
\begin{equation}
\varphi_1^*(\vec{R}_j) \left[\varphi_1(\vec{R}_j),\varphi_2(\vec{R}_j),\varphi_3(\vec{R}_j), \dots\right]
\end{equation}
 is proportional to the second row 
\begin{equation}
\varphi_2^*(\vec{R}_j) \left[\varphi_1(\vec{R}_j),\varphi_2(\vec{R}_j), \varphi_3(\vec{R}_j), \dots\right]
\end{equation}
etc. Consequently, the rank of the matrix representation (\ref{eqn:delta_appendix}) is maximally one and there is, hence,
maximally one eigenstate of this matrix with a non-zero eigenvalue. This is the trilobite state (\ref{eqn:trilobite}) 
\begin{equation}
\ket{\phi_j}= \sum_\alpha \varphi_\alpha^*(\vec{R}_j) \ket{\varphi_\alpha} 
\end{equation}
with eigenvalue $\displaystyle \sum_\alpha |\varphi_\alpha(\vec{R}_j)|^2 =\braket{\phi_j|\phi_j}$.
For this reason one can replace the delta potential (in the here considered basis set) by
\begin{equation}
\delta(\vec{r}-\vec{R}_j)= \ket{\phi_j} \bra{\phi_j} , 
\end{equation}
which proofs (\ref{eqn:pseudo_appendix}).

\end{document}

%% file: main.bbl
%